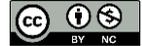

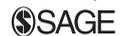

# Optical verification and in-vitro characterization of two commercially available acoustic bubble counters for cardiopulmonary bypass systems

Tim Segers,[1]\ Marco C. Stehouwer,[2]\ Filip M.J.J. de Somer,[3]
Bastian A. de Mol[4] and Michel Versluis[1]

## Abstract

*Introduction:* Gaseous microemboli (GME) introduced during cardiac surgery are considered as a potential source of morbidity, which has driven the development of the first bubble counters. Two new generation bubble counters, introduced in the early 2000s, claim correct sizing and counting of GME. This in-vitro study aims to validate the accuracy of two bubble counters using monodisperse bubbles in a highly controlled setting at low GME concentrations.

*Methods:* Monodisperse GME with a radius of 43 μm were produced in a microfluidic chip. Directly after their formation, they were injected one-by-one into the BCC200 and the EDAC sensors. GME size and count, measured with the bubble counters, were optically verified using high-speed imaging.

*Results:* During best-case scenarios or low GME concentrations of GME with a size of 43 μm in radius in an *in-vitro* setup, the BCC200 overestimates GME size by a factor of 2 to 3 while the EDAC underestimates the average GME size by at least a factor of two. The BCC200 overestimates the GME concentration by approximately 20% while the EDAC overestimates the concentration by nearly one order of magnitude. Nevertheless, the calculated total GME volume is only over-predicted by a factor 2 since the EDAC underestimates the actual GME size. For the BCC200, the total GME volume was over-predicted by 25 times due to the over-estimation of GME size.

*Conclusions:* The measured errors in the absolute sizing/counting of GME do not imply that all results obtained using the bubble counters are insignificant or invalid. A relative change in bubble size or bubble concentration can accurately be measured. However, care must be taken in the interpretation of the results and their absolute values. Moreover, the devices cannot be used interchangeably when reporting GME activity. Nevertheless, both devices can be used to study the relative air removal characteristics of CPB components or for the quantitative monitoring of GME production during CPB interventions.

## Keywords

GME; cardiopulmonary bypass; bubble counter; microbubble; emboli; validation

## Introduction

Since the early days of cardiac surgery, gaseous microemboli (GME) have been considered as a potential source of morbidity.[1–5] Major causes of GME are inefficient de-airing of the heart cavities and the use of cardiopulmonary bypass (CPB) systems.[1] To monitor GME and to study their adverse effects, bubble counters were developed. The working principle of all clinical bubble counters is based on ultrasound techniques, which had already been explored for decades, with particular interest in bubble counting and sizing in liquids.[6,7] Since the early 2000s, two new-generation clinical bubble counters have become available, with a higher accuracy than their predecessors: the BCC200 (Gampt, Zappendorf,

[1]Physics of Fluids Group, MESA+ Institute for Nanotechnology and MIRA Institute for Biomedical Technology and Technical Medicine, University of Twente, Enschede, The Netherlands
[2]Department of Extracorporeal Circulation, St Antonius Hospital, Utrecht, The Netherlands
[3]Heart Centre, Ghent University, Ghent, Belgium
[4]Section of Cardiovascular Biomechanics, Faculty of Biomedical Technology, Technical University Eindhoven, The Netherlands

\Equal contributions

**Corresponding author:**
Tim Segers, Physics of Fluids Group, MESA+ Institute for Nanotechnology, and MIRA Institute for Biomedical Technology and Technical Medicine, University of Twente, Enschede, The Netherlands.
Email: timsegers1@hotmail.com

Germany) and the EDAC (Luna Innovations, Roanoke, VA, USA). The manufacturers claimed quantitative sizing and counting of both small and large concentrations of GME. Many studies have been conducted using both bubble counters, evaluating the air removal properties of CPB circuits and their components.[8–13]

Validation of the bubble counters is a challenge as a variety of complex and specific techniques are needed to generate GME and for the control of GME properties, i.e. size, position, velocity and concentration. Several studies were conducted using controlled methods to produce microbubbles for the validation of bubble counters, e.g., using a pressurized micropipette in an outer liquid flow.[14–16] The two new generation bubble counters have been validated independently.[14,17–19] However, only one paper studied the accuracy of these devices in a one-to-one comparison in the same in-vitro setup.[20] In this study, GME were introduced by pumping a fluid through a partially air-filled arterial filter. By increasing the flow, the GME concentration was strongly increased; however, the GME size was also changed. Moreover, very high concentrations of GME were produced at or above the detection limit of the bubble counters.[21] It has been found that, under these so-called worst-case scenarios, both bubble counters only partially count the number of bubbles. Moreover, the BCC200 overestimated the average GME size whereas the EDAC underestimated the GME size.

This in-vitro study aims to validate the accuracy of two bubble counters at low GME concentrations in a highly controlled setting. The sizing and counting accuracies were measured during this so-called best-case scenario using bubbles of exactly the same size produced with a novel microfluidic technique. The monodispersity of the bubbles allowed for a direct one-to-one comparison of the injected bubble size to the bubble size distribution measured by the bubble counters. Moreover, the microfluidic production method facilitated optical sizing and optical counting of each single bubble, allowing for number count verification. In the first setup, the bubble counters were validated at a total flow-rate of 4 L/min through standard 3/8 inch CPB tubing. In the second setup, the bubbles produced were guided one-by-one, at 10 different locations, through the sensors of the bubble counters with the aim of gaining a fundamental understanding of the counting and sizing characteristics to allow for future design improvements, which should lead to a more accurate bubble counter. This work, therefore, aims to provide perfusionists with background knowledge on the accuracy of acoustic bubble sizing and counting.

## Methods

The sizing and counting characteristics of two contemporary bubble counters were studied; the BCC200 and the EDAC. Both bubble counters consist of two sensors wired to a base unit. The sensors need to be mounted on the CPB tubing.

### Working principle of the bubble counters

The working principle of acoustic bubble counters will now be briefly discussed. Bubble counters operate at an ultrasound frequency much higher than the resonance frequency[7] of the bubble sizes they aim to measure. In this regime, the incident ultrasound wave is geometrically scattered by the bubble and its pressure amplitude $P_s$ can be expressed as:[22]

$$P_s(r) = \frac{R}{r} P_a(R) \qquad (1)$$

Thus, there is a linear dependence of the amplitude of the scattered pressure wave[7] $P_s$, at a distance $r$ from the bubble, on the bubble radius $R$. Moreover, Equation (1) shows that the bubble size $R$ can only be determined when the distance between the bubble and the receiving element $r$ is known and when the transmit pressure field is well calibrated in order for the pressure amplitude to be known[22] at the bubble wall $P_a(R)$.

Both, the BCC200 and the EDAC use a technique called pulsed-Doppler. A pulser-receiver circuit transmits short ultrasound pulses under a fixed angle θ with respect to the blood flow direction, using a single element transducer (Figure 1A). Between each transmitted pulse, the piezo transducer is used as a receiver to record the bubble echo (Figure 1B). A time window is used in which several pulses are transmitted and recorded. When a single bubble passes through the ultrasound field within this time window, the amplitude of the resulting echo signal first increases and then decreases (Figure 1B). Echoes from non-moving objects have a constant echo amplitude, therefore, it is used to discriminate moving from non-moving scatterers.

The BCC200 only measures the largest bubble echo in the echo envelope within a time window (Figure 1). In contrast, the EDAC system uses a tracking algorithm to track all moving echoes within the time window and it calculates a bubble size from all local envelope maxima within the time window[19,20,29] ($A_1$ and $A_2$ in Figure 1B). As a result, the bubble echo (red) reflected from the wall (blue) will also be sized and counted as if it were a bubble.

### Setup 1

First, the bubble counters were validated in a simple setup with a constant flow through standard 3/8 inch CPB tubing. Water was pumped from a 10 L reservoir at a flow rate of 4 L/min through a loop of 3/8 inch CPB tubing (Figure 2A). An arterial filter (Maquet, Hirrlingen, Germany) was placed in the loop to remove bubbles from the flow. Further downstream, a square $1 \times 1$ cm² optically transparent tube was placed in the CPB tubing to create an optically transparent window, allowing for optical sizing of the injected bubbles (Figure 2B). The bubbles were monodispersed and they were produced one-by-one at a production rate of approximately 10 Hz in a microfluidic flow-focusing[24] chip, made out of polydimethylsiloxane (PDMS) using standard soft lithography techniques[25] (Figure 2C). The formed bubbles were guided from the chip to the CPB circuit by a capillary (Figure 2B; fused silica, I.D. 150 μm, CM Scientific, Silsden, West Yorkshire, UK). The bubble size was fixed at a size of 43 μm in radius since this was reported to be a common GME size during CPB.[26] Every injected bubble was optically imaged using a high-speed camera (Photron SA1.1 operated at 125 frames per s, Acal BFi Netherlands BV, Eindhoven, The Netherlands) connected to a 10 x magnifying objective lens (Olympus, LMPlanFLN). The sensors of the bubble counters were clamped onto the CPB tubing.

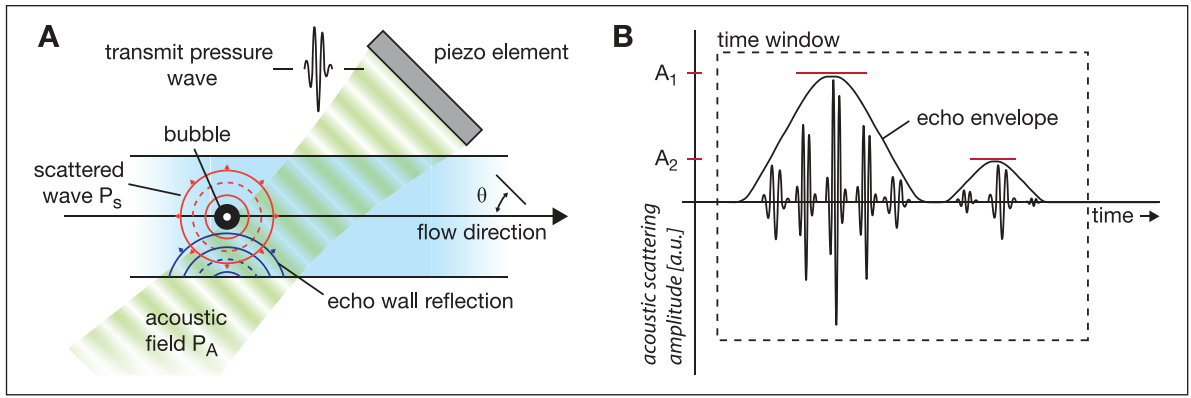

**Figure 1.** (A) Schematic representation of the pulsed-Doppler technique. A piezo element transmits ultrasound pulses under a fixed angle θ with respect to the blood flow direction. In between the transmitted pulses, the scattered pressure waves originating from a bubble passing through the acoustic field are recorded by the same piezo element. (B) The amplitudes of the recorded scattered waves first increase then decrease when a bubble passes through the acoustic field. A single bubble passing through the acoustic field produces a primary echo with amplitude $A_1$. It may also produce secondary echoes of lower amplitude $A_2$ due to reflections of the primary echo from the walls of the bubble counter probe.

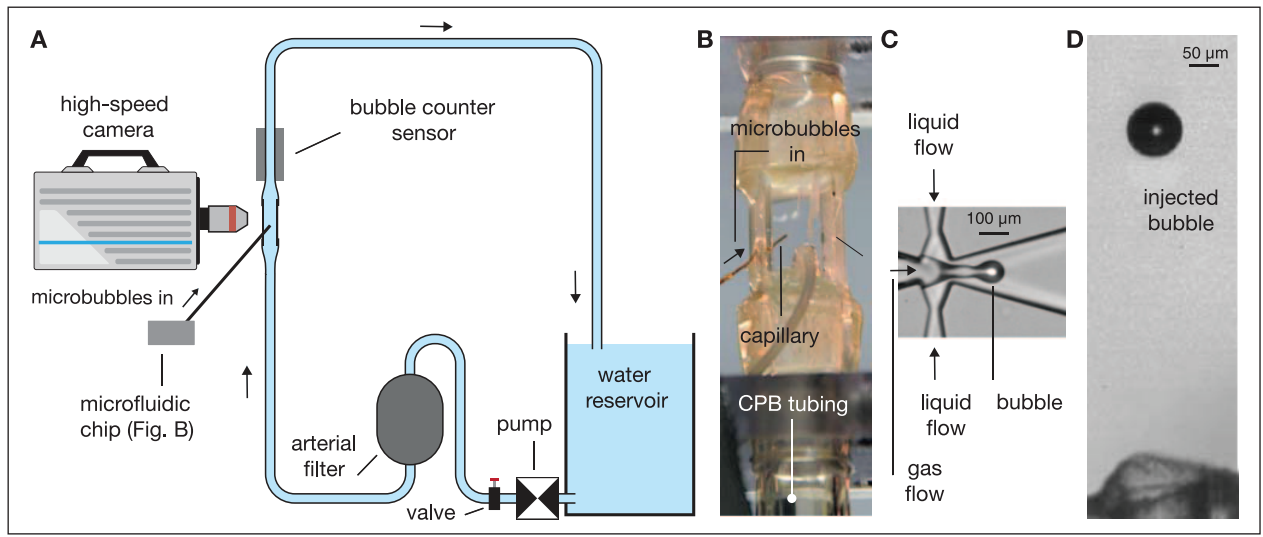

**Figure 2.** (A) Schematic representation of setup 1. Bubbles produced in the microfluidic device were directly injected into a 4 L/min flow of water through a 3/8 inch CPB tube and, subsequently, detected by the bubble counter sensor. (B) Optical window at the bubble injection location for optical verification using a high-speed camera. (C) Micrograph of the microfluidic bubble production device. (D) Every single injected bubble was optically verified.

Each measurement was conducted by simultaneously starting an optical recording and a bubble counter measurement. Then, the gas flow to the flow-focusing device was opened to start the injection of microbubbles. After 2.5 minutes, the bubble production was stopped and the optical measurement and bubble counter measurement were stopped. Each recorded movie contained 18,750 frames and these were processed frame by frame with an automated image-processing algorithm programmed in Matlab.[27] From each movie frame, the software detected whether a bubble was present and it measured its size (Figure 2D), thereby, providing the optical verification, i.e., the bubble size and the bubble production rate.

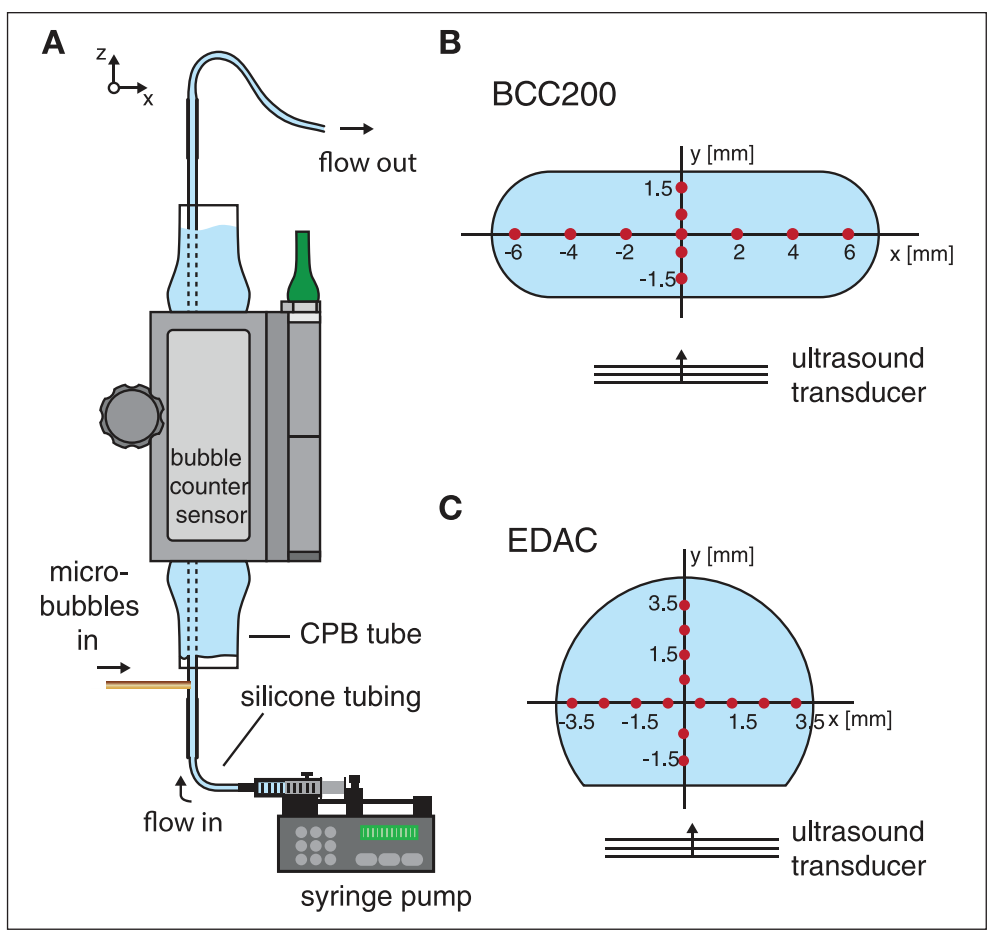

**Figure 3.** (A) Bubbles were guided by a capillary into an acoustically transparent silicone tube that guided the bubbles through the detectors at fixed locations. (B) Cross-section of a 0.95-cm diameter CPB tube mounted in the BCC200 sensor. (C) Cross-section of the CPB tube mounted in the EDAC sensor. Measurements were performed with the silicone tube at the locations marked by the red dots.

### Setup 2

Equation 1 shows that it is important to know the distance between the ultrasound transducer and the bubble. Therefore, in the second setup, the bubble position within the CPB tube was fixed and varied over 11 different locations in order to measure the effect of the bubble position on the counting and the sizing performance of the bubble counters. The bubbles were produced exactly as before; in the microfluidic chip. The microfluidic chip was optically transparent (Figure 2C), allowing for optical sizing of the bubbles. Every single microbubble produced was imaged within the chip, using an inverted microscope (Eclipse TE2000-U, Nikon Instruments,

Amsterdam, the Netherlands) equipped with a 6X objective lens, connected to the high-speed camera (Photron SA1.1 operated at 125 frames per s). The optical resolution was 3 µm/pixel, resulting in well-resolved optical sizing of the produced bubbles.

The formed bubbles were injected into an acoustically transparent silicone tube (I.D. 500 µm) (Figure 3A). The location of the silicone tube inside the bubble counter sensors could be varied in order to vary the location of the bubbles inside the sensors. The flow rate in the silicone tube was controlled using a syringe pump (Harvard Apparatus, PHD2000, Cambourne, Cambridgeshire, UK) at a total flow-rate of 6.6 mL/min. This resulted in a translational bubble velocity equal to that of a flow with a flow rate of 4.8 L/min through standard clinical 3/8-inch diameter CPB tubing.

The BCC200 sensor was positioned around rubber-sealed 3/8-inch tubing (Figure 3A) and filled with water for acoustic coupling. The silicone tube was positioned inside the 3/8 inch tubing parallel to the tubing wall (Figure 3A). The silicone tube was fixed by a U-shaped holder mounted on an x-y translation stage to vary its position in the sensor of the bubble counter. The EDAC sensor was mounted in a similar way except that the sensor had to be mounted on a disposable plastic cuvette (Luna Innovations, Roanoke, VA, USA). The cross-section of the cuvette is different compared to that of the BCC200 sensor. The red dots in Figures 3B and 3C represent the 11 locations (BCC200) and the 14 locations (EDAC) at which data were collected. The measurement procedure was as described in the previous section.

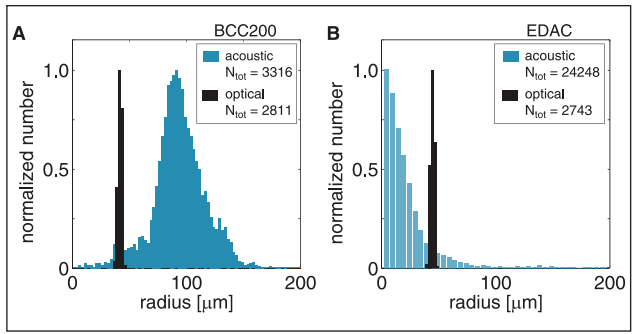

**Figure 4.** (A) The injected normalized bubble size distribution (black) and the normalized size distribution measured with the BCC200 (blue) measured with setup 1. The total acoustic number count was 3316 and the total optical number count was 2811. (B) Repeated measurement with the EDAC. The total acoustic number count was 24,248 and the total optical number count was 2743.

## Results

### Setup 1

*BCC200.* Figure 4A shows the injected bubble size distribution (black) and the size distribution measured with the BCC200 (blue). The bubble size is over-predicted by a factor of 2. The total number count is shown in the legend of Figure 4A and it was 1.2 times higher than the total number of injected bubbles.

*EDAC.* For the EDAC system, the injected and the acoustically measured bubble size distributions are shown in Figure 4B. The EDAC underestimates the bubble size and large numbers of small bubbles were counted (blue) whereas monodisperse bubbles were injected (black). Please note that the number counts are normalized to one. The total number count measured with the EDAC was 8.8 times higher than the optical number count.

### Setup 2

*BCC200.* A typical measurement performed with the BCC200 system is presented in Figures 5A and 5B. Figure 5A shows the optically measured size distribution (black) and the acoustically measured size distribution (blue) of the BCC200 system at position [0,0] in Figure 3B. A relatively narrow acoustically measured size distribution is observed (R = 90 ± 22 μm); however, the absolute bubble size is overestimated by a factor of 2, as can be observed from the optical verification measurement (43 ± 0.9 μm). The optical (black dots) bubble count and the acoustical (blue dots) bubble count over time are shown in Figure 5B. The total acoustic number count follows the trend of the optical verification. However, it is 1.2 times higher.

The measurements were repeated at all positions marked in Figure 3B with the BCC200 system (Figures 5C and 5D). The gray areas represent the acoustically measured size distribution, the black dots the optically measured bubble size and the mode of the acoustic bubble counts (bubble radius with largest count number) is represented by the blue horizontal lines. Since the size distributions are not symmetrically distributed around the mode, the standard deviation of the data on both sides of the mode are shown (black error bars). Please note that the data presented in Figure 5A are also presented in Figure 5C at [0,0], see arrow. At two measurement positions ([−4,0] and [4,0]), the system measures a bimodal size distribution with two local maxima. Figures 5E and 5F show the acoustic bubble counts normalized by the optically measured bubble counts at each measurement location.

From the size distributions and the bubble counts, the total injected gas volume and acoustically measured gas volume were determined and they were 8.3 μL and 198 μL, respectively. While the BCC200 system accurately counted bubbles, it overestimated the total injected gas volume by a factor 25 due to the over-prediction of the bubble sizes by a factor 2 to 3 (volume ∝ $R^3$).

*EDAC.* A typical measurement performed with the EDAC system at position [0,0.5] is shown in Figures 6A and 6B. The acoustically measured size distribution shows a high number of counts at bubble sizes down to the smallest detectable bubble size by the EDAC system (2.5 μm in radius). Thus, the EDAC system largely overestimates the amount of bubbles passing its probe. This is confirmed by the number count shown in Figure 6B, it shows that the EDAC system overestimated the bubble passing-rate on average by 9-10 times at this measurement location for the injected bubbles with a radius of 43 μm. The size distributions measured with the EDAC system at the measurement locations are shown in Figures 6C and 6D. Note that the acoustically measured size distributions have a local maximum positioned close to the optically measured bubble size (black dots). All acoustically measured size distributions show large number counts at bubble radii down to the smallest detectable bubble size. No standard deviations were calculated because of the non-Gaussian nature of the acoustically measured size distributions. The ratio of the total acoustic bubble count to the total optical bubble count at every measurement position is shown in Figures 6E and 6F. On average, the number of bubbles passing the probe is overestimated by a factor 9 for the injected bubble radius of 43 μm. At bubble positions close to the center of the channel and close to the wall opposing the transducer, see Figure 3C, the mismatch between the acoustic bubble count and the optical bubble count is the largest. The total injected gas volume and the total measured gas volume were 7.2 μL and 13.8 μL, respectively, thus, with an overprediction approximately by a factor 2.

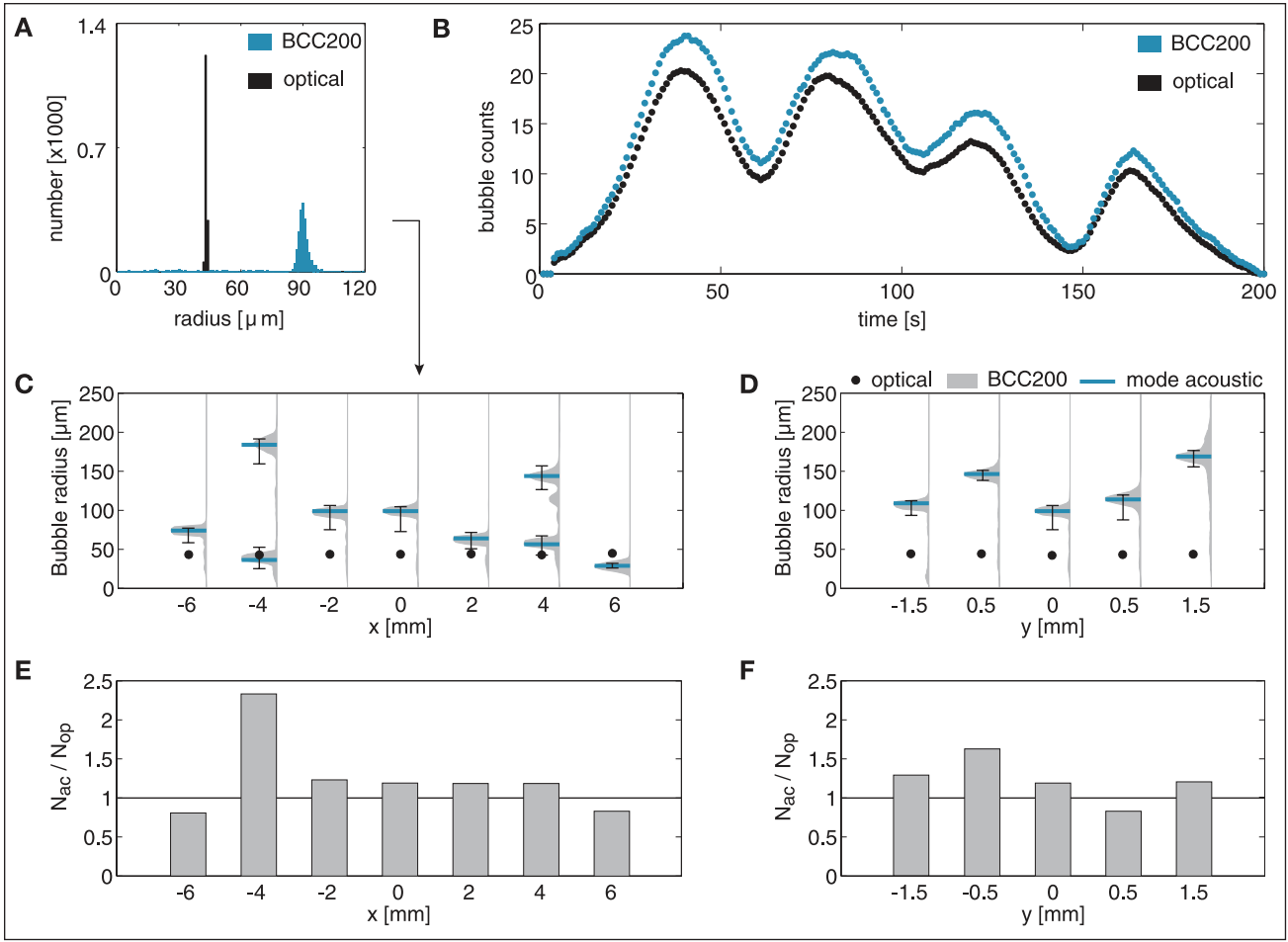

**Figure 5.** (A) Optically and acoustically measured bubble size distributions with the BCC200 system at position [0,0] in Figure 3B. (B) Bubble counts as a function of time measured with the BCC200 system. (C) Acoustically measured size distribution (gray areas) as a function of the horizontal (C) and vertical (D) bubble position in the BCC200 system. Blue horizontal lines represent the mode of the size distribution. The standard deviation is represented by the vertical black dashes. Optically measured bubble radius is represented by the black dots. Figures E and F show the ratio of the acoustic to the optical bubble count for horizontal and vertical bubble positions, respectively.

## Discussion

The total number count measured using setup 1 and that measured using setup 2 are in excellent agreement. The BCC200 overestimated the total number count by a factor of 1.2, independent of the employed setup and the total number count measured using the EDAC was overestimated by a factor of 8.8 in setup 1 and 9 times in setup 2. On top of that, the BCC200 overestimated the bubble size by a factor of 2, also setup independent. Therefore, it can be concluded that the acoustically transparent silicon tube that fixed the bubble position in setup 2 did not significantly influence the number count and size measurements performed using setup 2.

The BCC200 counts GME with reasonable agreement since it measures only the largest bubble echo (Figure 1). In contrast, the EDAC overestimates the number count by almost 9 times during the presented measurements at relatively low GME concentrations. This overestimation is likely due to the fact that the EDAC system calculates a bubble size from all echoes; bubble echoes as well as bubble echoes reflected from the interfaces of the EDAC sensor and its disposable plastic cuvette.

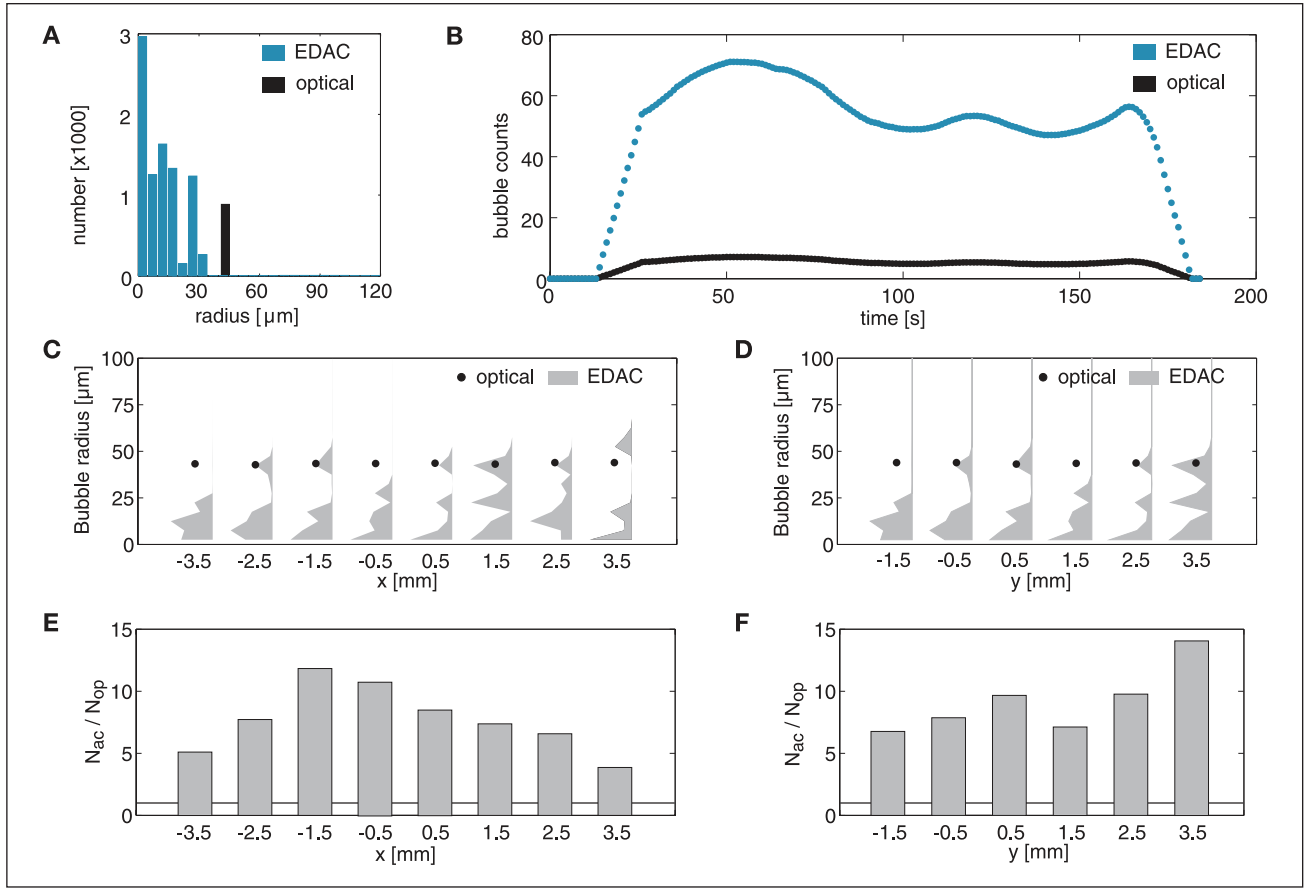

**Figure 6.** (A) Optically and acoustically measured size distributions with the EDAC system at position [0,0.5] in Figure 3C. (B) Bubble counts as a function of time measured with the EDAC system. Acoustically measured size distribution (gray areas) as a function of the horizontal (C) and vertical (D) bubble position in the EDAC system. The black horizontal lines represent the mode of the size distribution of which the standard deviation on both sides is shown by the vertical black error bars. The optically measured bubble radius is represented by the black dots. Figures E and F show the acoustic to optical bubble count ratio for horizontal and vertical bubble positions, respectively.

A reason for incorrect bubble sizing could be an incorrect measurement or the total omission of the distance between the bubble and the transducer (see Equation 1). The BCC200 system does not account for the exact position of the bubble.[18] It is not known if the EDAC system accounts for the exact position of the bubble as we have no direct access to the sizing algorithms in the software. Nevertheless, no decrease in acoustically measured bubble size is observed in Figures 5D and 6D for increasing bubble-to-transducer distances, i.e., for increasing y-values, which is expected since the bubble echo amplitude decreases with $1/r$. What is expected to be of larger influence here is that the acoustic pressure field $P_a$, generated by a single element transducer, is not uniform,[19,28] an artifact that is not accounted for in the sizing software. The consequence of a non-uniform acoustic pressure amplitude distribution is that the acoustic pressure at the bubble position is unknown if the position of the bubble in the acoustic field is unknown. This leads to a substantial error in the calculated bubble size. On top of that, the sensitivity of the piezo transducer to acoustic waves from point sources located away from the natural focus[28] of the transducer, i.e., echoes originating from bubbles close to the boundaries of the measurement volume, decreases with increasing lateral distances with respect to the natural focus. This leads to another additive error in the bubble size measurement.

The size distribution in Figure 4A is relatively narrow compared to the average size distribution of Figures 5C and 5D. This is due to the fact that the bubbles are injected in the center of the PCB tube, thereby, passing through the bubble counter sensor at a relatively narrow spatial distribution.

The authors only have knowledge about the validation process of the BCC200. This device is calibrated using an internal protocol wherein microbubbles are produced and optically and acoustically measured. However, since the technique of controlled microbubble generation is not straightforward, this calibration process might be sensitive to errors and a correct calibration may lead to a higher accuracy. Nevertheless, the over-prediction of the bubble size by a factor of 2 seems to be device independent since it was also found by de Somer et al.[20]

Water at room temperature was used during the measurements instead of whole blood. The attenuation of ultrasound in blood is 0.2 dB/cm/MHz,[29] which is approximately 100 times higher than the attenuation coefficient of water.[29] The amplitude of acoustic waves of 2 MHz (BCC200) and 4 MHz (EDAC) traveling through blood will have a 3% and 6% lower amplitude, respectively, for every 1 cm of propagation distance, compared to the same wave that travels through water. Because of the linear relationship between the acoustically measured bubble size and the pressure amplitude, the effect of water over blood is considered to be negligible in comparison to the observed sizing errors. The temperature of the medium does affect its viscosity and it may affect bubble stabilization, however, it does not affect the geometrical scattering of a gas bubble (see Equation 1). Therefore, the sizing and counting performance of the bubble counters is expected to be temperature independent. Nevertheless, further research is required in this direction, in the direction of water versus blood and in the role of the flow rate before these results can be translated to a clinical setting.

Monodisperse bubbles of a single size, namely 43 μm in radius, were used here and the sizing and counting performances were not examined for other bubble sizes as these measurements already clearly reveal the limitations of both bubble counters regarding their sizing and counting capabilities. Moreover, our results regarding the sizing and counting errors are in good agreement with those presented by de Somer et al.[20] who used a suspension of polydisperse bubbles. Different bubble sizes are not expected to have a different outcome on the sizing and counting performances of the BCC200 system since only the largest bubble echo is measured. This is different for the EDAC system. It is expected that bubble sizes larger than those examined here will result in an overestimation of the bubble concentration to an even higher degree because the larger bubbles will produce larger primary echoes and, as a result, more secondary echoes (reflections) will be detected above the noise level.

Finally, it is of interest to discuss the implications of the present work. The measured errors in the absolute sizing and counting of GME presented here do not imply that all previous work obtained using the bubble counters are insignificant or invalid. A relative change in bubble size and/or bubble concentration can readily be measured and, therefore, they can be used to study relative air removal characteristics of CPB components and/ or bubble activity in-vivo. However, care must be taken in the interpretation of the absolute numbers. Therefore, the development of an accurate bubble counter would benefit clinical research on the correlation between negative neurocognitive outcome after CPB and quantitative GME size and concentration.

## Conclusions

During a best-case scenario at low bubble concentrations of bubbles with a size of 43 μm in radius in an in-vitro setup, the BCC200 overestimates GME size by a factor of 2 to 3 while the EDAC underestimates the average bubble size by at least a factor of two. The BCC200 overestimates the bubble concentration by approximately 20% while the EDAC overestimates the concentration by nearly one order of magnitude. Nevertheless, the calculated total gas volume is only over-predicted by a factor 2 since the EDAC underestimates the actual bubble size. For the BCC200, the total gas volume was over-predicted up to 25 times due to the overestimation of the bubble size. The measured errors in the absolute sizing/counting of GME presented here do not imply that all results obtained using the bubbles counters are insignificant or invalid. A relative change in bubble size and/or bubble concentration can readily be measured and, therefore, they can be used to study relative air removal characteristics of CPB components and/or bubble activity in-vivo. However, care must be taken in the interpretation of the results and their absolute values. Moreover, both devices cannot be used interchangeably when reporting GME activity.

### Acknowledgements

We thank Gampt GmbH for kindly responding to all our questions. We also want to thank Gert-Wim Bruggert, Martin Bos and Bas Benschop for their skillful technical assistance.

### Declaration of Conflicting Interests

The authors declared no potential conflicts of interest with respect to the research, authorship, and/or publication of this article.

### Funding

The authors disclosed receipt of the following financial support for the research, authorship, and/or publication of this article: This work is supported by NanoNextNL, a micro and nanotechnology consortium of the Government of the Netherlands and 130 partners.

## References

1. Barak M, Katz Y. Microbubbles: pathophysiology and clinical implications. *Chest* 2005; 128: 2918–2932.
2. Rodriguez RA, Williams KA, Babaev A, Rubens F, Nathan HJ. Effect of perfusionist technique on cerebral embolization during cardiopulmonary bypass. *Perfusion* 2005; 20: 3–10.

3.  Barbut D, Lo Y, Gold J, et al. Impact of embolization during coronary artery bypass grafting on outcome and length of stay. *Ann Thorac Surgery* 1997; 63: 998–1002.

4.  Clark RE, Brillman J, Davis DA, Lovell MR, Price TRP, Macgovern GL. Microemboli during coronary artery bypass grafting. *J Thorac Cardiovasc Surg* 1995; 109: 249–258.

5.  Raymond PD, Marsh NA. Alteration to haemostasis following cardiopulmonary bypass and the relationship of these changes to neurocognitive morbidity. *Blood Coag Fibrinolysis* 2001; 12: 601–618.

6.  Nishi RY. Ultrasonic detection of bubbles with Doppler flow transducers. *Ultrasonics* 1972; 10: 173–179.

7.  Medwin H. Counting bubbles acoustically: a review. *Ultrasonics* 1977; 15: 7–13.

8.  De Somer F. Impact of oxygenator characteristics on its capability to remove gaseous microemboli. *J Extra Corpor Technol* 2007; 39: 271–273.

9.  Stehouwer MC, Boers C, de Vroege R, Kelder C, Yilmaz A, Bruins P. Clinical evaluation of the air removal characteristics of an oxygenator with integrated arterial filter in a minimized extracorporeal circuit. *J Artif Organs* 2011; 34: 374–382.

10. Liu S, Newland RF, Tully PJ, Tuble SC, Baker RA. In vitro evaluation of gaseous microemboli handling of cardiopulmonary bypass circuits with and without integrated arterial line filters. *J Extra Corpor Technol* 2011; 43: 107–114.

11. Potger KC, McMillan D, Ambrose M. Microbubble transmission during cardiotomy infusion of a hardshell venous reservoir with integrated cardiotomy versus a softshell venous reservoir with separated cardiotomy: an in vitro comparison. *J Extra Corpor Technol* 2013; 45, 77–85.

12. Johagen D, Appelblad M, Svenmarker S. Can the oxygenator screen filter reduce gaseous microemboli. *J Extra Corpor Technol* 2012; 46: 60–66.

13. Melchior RW, Rosenthal T, Glatz AC. An in vitro comparison of the ability of three commonly used pediatric cardiopulmonary bypass circuits to filter gaseous microemboli. *Perfusion* 2010; 25: 255–263.

14. Lynch JE, Pouch A, Sanders R, Hinders M, Rudd K, Sevick J. Gaseous microemboli sizing in extracorporeal circuits using ultrasound backscatter. *Ultrasound Med Biol* 2007; 33: 1661–1675.

15. Palanchon P, Klein J, de Jong N. Production of standardized air bubbles: application to embolism studies. *Rev Sci Instrum* 2003; 74: 2558 –2563.

16. Banahan C, Hague JP, Evans DH, Patel R, Ramnarine KV, Chung EML. Sizing gaseous emboli using Doppler embolic signal intensity. *Ultraound Med Biol* 2012; 38: 824–833.

17. Lynch JB, Riley JE. Microemboli detection on extracorporeal bypass circuits. *Perfusion* 2008; 23: 23–32.

18. Jenderka KV, Schultz M, Dietrich G, Fornara P. Detection and sizing of micro bubbles in streaming fluids with ultrasound. *Sensors Proceedings of IEEE* 2002; 1: 528–531.

19. Brand S, Klaua R, Dietrich G, Schultz M. Ultrasonic detection and quantitative analysis of microscopic bubbles and particles in solution: enhanced by attenuation compensation. *Proceedings of the IEEE Ultrasonics Symposium* 2006; 1: 1840–1843.

20. De Somer FM, Vetrano MR, Van Beeck JP, van Nooten GJ. Extracorporeal bubbles: a word of caution. *Interact Cardiovasc Thorac Surg* 2010; 10: 995–1002.

21. Schultz M, Klaua R, Oblonczek G. eComment: Ultrasonic bubble detection: some additional information. *Interact Cardiovasc Thorac Surg* 2010; 10: 1001–1002.

22. Maresca D, Emmer M, van Meer PLMJ, et al. Acoustic sizing of an ultrasound contrast agent. *Ultrasound Med Biol* 2010; 36: 1713–1721.

23. Urbanek S, Tiedtke HJ. Improved methods for measurement of gaseous microbubbles during extracorporeal circulation. *Perfusion* 2012; 17: 429–434.

24. Tan YC, Cristini V, Lee AP. Monodispersed microfluidic droplet generation by shear focusing microfluidic device. *Sensors and Actuators B: Chemical* 2005; 114: 350–356.

25. Duffy DC, McDonald JC, Schueller OJA, Whitesides GM. Rapid prototyping of microfluidic systems in poly(dimethylsiloxane). *Anal Chem* 1998; 70: 211–217.

26. Chung EM, Banahan C, Patel N, et al. Size distribution of air bubbles entering the brain during cardiac surgery. *PLoS One* 2015; 10: e0122166.

27. Segers T, Versluis M. Acoustic bubble sorting for ultrasound contrast agent enrichment. *Lab Chip* 2014; 14: 1705–1714.

28. Szabo T. *Diagnostic Ultrasound Imaging: Inside Out*. London: Academic Press, 2004.

29. Culjat MO, Goldenberg D, Tewari P, Singh R. A review of tissue substitutes for ultrasound imaging. *Ultrasound Med Biol* 2010; 36: 861–873.